\documentclass[12pt,preprint]{aastex}

\newcommand{\myemail}{lcz01@mails.tsinghua.edu.cn}

\shorttitle{Spectral Variability in Cyg X-1}
\shortauthors{Liu \& Li}

\begin{document}
\title{X-RAY SPECTRAL VARIABILITY IN CYGNUS X-1}
\author{C.Z. Liu\altaffilmark{1} and T.P. Li\altaffilmark{1,~ 2}}
\altaffiltext{1}{Department of Engineering Physics \& Center for
Astrophysics, Tsinghua University, Beijing, China;
\myemail}
\altaffiltext{2}{Particle Astrophys.
Lab., Inst. of High Energy Phys., Chinese Academy of Sciences,
Beijing, China}

\begin{abstract}
Spectral variability in different energy bands of X-rays from
Cyg X-1 in different states is studied with $RXTE$ observations
and time domain approaches. In the hard tail of energy spectrum
above $\sim 10$\,keV, average peak aligned shots are softer than
the average steady emission and the hardness ratio decreases
when the flux increases during a shot for all states. In regard
to a soft band lower $\sim 10$\,keV, the hardness in the soft
state varies in an opposite way: it peaks when the flux of the
shot peaks. For the hard and transition states, the hardness
ratio in respect to a soft band during a shot is in general
lower than that of the steady component and a sharp rise is
observed at about the shot peak. For the soft state, the
correlation coefficient between the intensity and hardness ratio
in the hard tail is negative and decreases monotonically as the
timescale increases from 0.01\,s to 50\,s, which is opposite to that
in regard to a soft band. For the hard and transition states,
the correlation coefficients are in general negative and have a
trend of decrease with increasing timescale.
\end{abstract}


\section{INTRODUCTION}
 The intensity and spectral variabilities of the high energy
emission from X-ray binaries carry valuable information about
their production regions and mechanisms around compact objects.
Cyg X-1, one of the most early discovered and best-studied black
hole X-ray binaries \citep{Bow65}, is the brightest source in
the hard X-ray sky suitable for the study of rapid variability.
The compact object has a mass of $\sim$$10$\,$M_{\odot}$ and the
companion star HDE226868 discovered
in an optical observation by Walborn (1973) has a mass of
 $\sim$18$M_{\odot}$ \citep{Her95}.
 The X-ray emission of Cyg X-1 is very complex in a wide
time scale from months to milliseconds. In a long-term period, Cyg
X-1 generally shows three different states: hard, soft and
transition states.

Many temporal and spectral properties
of X-rays from Cyg X-1 have been studied before, e.g., power
density spectra (PDS), time lags and coherences in different energy
bands and on different time scales. Most of these studies are
carried out in the frequency domain by using Fourier transformation.
Studies for different states of Cyg X-1 have been done with the
observation data of the Proportional Counter Array (PCA) on board the
{\sl Rossi X-ray Timing Explorer} ({\sl RXTE}), e.g.,
studies of temporal properties of Cyg X-1 in the soft state
\citep{cui97a}, during the spectral transitions \citep{cui97b},
and in the hard state \citep{now99}. The hard state PSD of Cyg X-1
in the Fourier frequency range between 0.001\,Hz and 100\,Hz can be
successfuly modeled  with multiple energy-dependent Lorentzians -- distinct
broad noise components (Nowak 2000; Pottschmidt et al. 2003).

A characteristic shape of a Fourier PDS or a structure in the PDS,
e.g. a quasi-periodic oscillation
(QPO) or a broad noise component, can be generated by different kinds
of process. For finally understanding the undergoing physical processes,
it is also necessary to study variabilities directly in the time domain.
On the subsecond timescale, the X-rays from Cyg X-1 show
large-amplitude chaotic fluctuations. Randomly occurring shots
may have significant contribution on the rapid variability (Terrel
1972; Oda 1977; Sutherland, Weisskopf \& Kahn 1978; Nolan et al.
1981; Meekings et al. 1984; Miyamoto et al. 1988, 1992; Lochner,
Swank \& Szymkowiak 1991).
To study the shot process
Negoro, Miyamoto \& Kitamoto (1994) constructed the average peak
aligned shot profile in the 1.2--58.4\,keV energy band and its
hardness ratios in (7.3--14.6\,keV)/(1.2--7.3\,keV) and (14.6--21.9\,keV)/(1.2--7.3\,keV) 
with $Ginga$ observation data of Cyg X-1 in
the hard state. They found that the shot becomes first softer,
and then, just after the peak, harder than the average emission.
With $RXTE$/PCA data of Cyg X-1 and an improved shot detection
algorithm, Feng, Li \& Chen (1999) studied evolution of hardness
ratios in (13--60\,keV)/(2--6\,keV) in the hard, soft and transition
states. Their results confirmed what Negoro, Miyamoto \&
Kitamoto (1994) had found in the hard state, and they found that
in the soft state, shots are harder, and in the transition
states softer than the time average emission. Li, Feng \& Chen
(1999) used the correlation analysis technique to study the
relationship between temporal and spectral variabilities in Cyg
X-1 in different states and the results are consistent with that
from the average shot analysis mentioned above.

The X-ray Spectrum of Cygnus X-1 can be decomposed into several
components : thermal
Comptonization, Compton reflection, and a soft excess.
We make use of $RXTE$/PSPC observations of Cyg X-1
to study spectral variability in different energy bands.
Poutanen (2001) in his review paper pointed out that
it would be of interest to see how the hardness in the energy
bands above $\sim 10$\,keV varies in the shots. This would
provide important information about spectral variability of the
hard (power-law like) tail.
We lay the emphasis on studying hardness evolution during shots and
the correlation coefficient between hardness and intensity in the hard tail
in this paper. Our results imply that the observed
behavior, and then the physical mechanism, of the hard component
is quite different from that of soft component.

\section{SPECTRAL EVOLUTION DURING SHOTS}
A series of observations of Cyg X-1 were performed by $RXTE$ in
1996. The All Sky Monitor on $RXTE$ revealed that Cyg X-1 started
a transition from the normal hard state to the soft state. It
stayed in soft state for nearly 3 months and then went back down
to hard state. The data used in our work are from the public
$RXTE$ archives, listed in the Table~\ref{tbl-1}. We extract the
PCA data with the version 4.2 of
standard $RXTE$ ftools for windows to get 1 ms time bin light
curves. In this process, the data were selected to use when the
source was observed at the elevation angle larger than
$10^{\circ}$, the offset pointing less than $0.02^{\circ}$, and
the number of PCUs turning ``ON'' equaling to 5. In hard
X-ray band, Cyg X-1 is a very bright source. The average of the 
background contribution for 13--60\,keV is at 10\% level and less 
than 1\% for lower bands (2--6\,keV \& 6--13\,keV). For the purpose of
studying the average features of shots, background can be negligible.
The $RXTE$ dead-time per event in one PCU is about 10\,$\mu$s \citep{Zhang99}. 
For Cyg X-1, the effect of dead-time is about 1\% (Maccarone, Coppi \&
Poutanen 2000), which is neglected in our analysis. We search shots 
from 1\,ms time bin light curves by using the algorithm proposed by 
Feng, Li \& Chen (1998), which is a modified algorithm to the peak 
detection technique of Negoro, Miyamoto \& Kitamoto (1994) and Li \& Fenimore (1996).
In order to suppress the effects of statistical fluctuation, a light curve
with 1\,ms time bin was first merged into larger time bin of 10\,ms.
Then the bin having more counts $C_p $ than the both sides
neighboring bins is selected as a candidate peak. In the
neighboring 1\,s on both sides of each candidate peak, we search
for the bins with counts $ C_1 $ and $ C_2 $ so that the condition
\begin{equation}  C_p > C_{1, 2} + \alpha\sqrt{C_p+C_{1, 2}}
\end{equation} is satisfied with $\alpha$ being selected from [2,
3]. If and only if the number of bins with counts $C_1$ and $C_2$
is larger than a certain criterion respectively, we affirm that
the count rate is significantly smaller around the candidate peak bin.
And then this candidate peak is selected as a potential shot peak. Then
each selected potential shot peak bin and its both sides
neighboring bins are divided into thirty bins with time bin of 1\,ms.
 A shot is finally selected by the criteria that their count
should be 2 times larger than that of the mean count of the
observation and should be the maximum within the thirty bins.

The shot detection process was performed respectively in each of
three energy bands 2--6\,keV, 6--13\,keV and 13--60\,keV (2--5\,keV, 
5--13\,keV and 13--60\,keV for the hard state). A shot peak is selected as
a true shot peak if it coincidences in all the three energy bands
within 30\,ms. We define the 400\,ms time interval neighboring a
shot peak as a shot period. For a studied pair of energy
bands $A$ and $B$, we calculate the evolution curve of hardness
ratios with $h = f_A/f_B$ in the shot period, where $f_A, f_B$ are
the count rate of band $A$ and $B$ respectively. Figure~1 shows
the total shot profile of 2--60\,keV energy band (dashed line) and
hardness variation (solid line) for different states and different
pairs of studied energy band. All the panels in Fig.~1 are
classified into four groups, a1-a4 for the soft state, b1-b4 for
the transition state of hard-to-soft, c1-c4 for the soft-to-hard
transition state, and d1-d4 for the hard state.

From the panels a4, b4, c4, and d4 of Fig.~1, one can see that the
hardness ratios in the power-law tail, (16--60\,keV)/(13--16\,keV),
vary negatively correlated with the intensity, the hardness
variation during a shot is dominated by a negative peak. This
feature is almost completely opposite to what is seen in the
hardness ratios in respect to an energy band below $\sim 10$\,keV
for the soft state: The hardness ratios of (6--13\,keV)/(2--6\,keV),
(13--60\,keV)/(2--6\,keV), and (13--60\,keV/(2--13\,keV) peak when the
corresponding fluxes of the shots peak, shown in a1, a2, and a3 in
Fig.~1. Maccarone \& Coppi (2002a) studied the short timescale 
correlations between line and continuum fluxes in Cyg X-1. They found
that the spectrum got harder as the source got brighter in soft state.
For the hard and transition states, the variation of hardness 
ratio in regard to a energy band below $\sim 10$\,keV behaves with 
a similar feature: the shot becomes first softer, and then, just 
around the peak, rapidly harder.

\section{CORRELATION BETWEEN HARDNESS AND INTENSITY}
We here study X-ray spectral variability of Cyg X-1 in different
energy bands and different timescales through correlation
analysis. Li, Feng \& Chen (1999) proposed an algorithm to
calculate the correlation coefficient $r(h,f)$ between the
hardness ratio $h$ and the total intensity $f$ on a given
timescale $\Delta t$. The total observation data of Cyg X-1 in a
spectral state is divided into $M$ periods. For an observation
period $k~(k = 1,2,...,M)$ and a given time bin $\Delta t$, the
effective data in two studied energy bands are divided into
segments with a duration of $n\Delta t,~ n=10$. The correlation
coefficient \begin{equation} r(h,f) =
\frac{\displaystyle\sum_{i=1}^n{(h(i)-\bar h)(f(i)-\bar
f)}}{\sqrt{\displaystyle\sum_{j=1}^n{(h(i)-\bar
h)^2}\displaystyle\sum_{i=1}^n{(f(i)-\bar f)^2}}} \end{equation}
is calculated for each segment and then their average $\bar r(k)$
and standard deviation $\sigma(\bar r(k))$ are derived. The
correlation coefficient on a timescale $\Delta t$ is estimated by
\begin{equation} \langle \bar r \rangle=\sum_{k=1}^M \omega(k)\bar
r(k)/\sum_{k=1}^M\omega(k) ,
\end{equation} where \ \begin{equation}\omega(k)=1/\sigma^2[\bar r(k)]
\end{equation} and its uncertainty is given by
\begin{equation} \sigma=\sqrt{\sigma_1^2+\sigma_2^2} ,\end{equation}
where
\begin{displaymath}\sigma_1^2=1/\sum_{k=1}^M\omega(k)
\end{displaymath}
\begin{equation} \sigma_2^2=\sum_{k=1}^M[\bar r(k)- \langle \bar
r \rangle]^2/M(M-1). \end{equation} Finally, the result for the
observation can be presented as : $\bar r$ = $\langle \bar r
\rangle$ $\pm ~\sigma$.            

Usually we can use some convenient statistical methods based on the 
normal distribution to make statistical inference, e.g. significance 
test, on $\bar r$. For the case of short time scale
$\Delta t$, although the number of counts per bin may be small for it 
to be assumed as a normal variable, it is easy to obtain a total number 
of segments from a certain observation that is large enough that the 
central limit theorem can be applied and that a normal distribution can 
be assumed for estimating the uncertainty of $\bar r$. 

For each state, we calculate the correlation coefficients
between intensity and hardness ratio of (13--60\,keV)/(2--6\,keV)
and (16--60\,keV)/(13--16\,keV) on time scales 0.01\,s, 0.1\,s, 1\,s,
10\,s, and 50\,s respectively. The results are shown in Table 2--3
and Figure 2.

From upper-left panel of Fig.~2 for the soft state, we can see that
the correlation coefficients between the intensity and hardness ratio
in the 2--6\,keV band are positive
on all timescales between 0.01\,s and 50\,s. In contract, they are negative
for the 13--16\,keV band. The two kinds of correlation vary
with timescale in opposite directions:
for the hardness ratios in regard to the
soft band, the correlation coefficient $r(h,f)$ increases
monotonically and reaches near unity as the timescale increases
from 0.01\,s to 50\,s; on the other hand, for $h$ in the
hard band, the correlation coefficient $r(h,f)$ decreases
monotonically and reaches near $-1$. The hard and transition
states have the similar feature in hardness-intensity correlation:
the correlation coefficients
are in general negative and have a trend of decrease with
increasing timescale.

Figure 3 shows example of light curve and hardness profiles in 50\,s
 time bin in the soft and hard states. From Fig.3 we can see
that, consistent with the results of correlation analysis, the
hardness ratios in (13--60\,keV)/(2--6\,keV) are positively related to
the light curve and the hardness in (16--60\,keV)/(13--16\,keV)
negatively for the soft state, and a rather complex mixture of
positive and negative correlations for the hard states.

\section{SUMMARY AND DISCUSSION}
The simplest single shot models assume that the observed
lightcurve consists of the superposition of individual uncorrelated
shots with a fixed profile and intensity. Lochner, Swank \& Szymkowiak
(1991) found that the simplest shot models cannot fit the overall
PSD shape of Cyg X-1 well. Recently Maccarone \& Coppi (2002b) found that
the single shot variability models also cannot produce the observed skewness
pattern. In this work we study the spectral evolution in different energy
bands of X-rays from Cyg X-1 in different states during the average peak
aligned shots in subsecond duration region, our results on shot are
in the average meaning, neither assuming the shots have a fixed profile
and nor assuming the shots can be responsible alone for the overall power spectrum.
The timescale analysis we performed for correlation between hardness and
intensity is not correlated with any special shot model.
Main features in spectral variability of Cyg X-1 X-rays
revealed by our time domain analysis can be summarized as follows:
(1) For any spectral state of Cyg X-1, on the average the hardness profile of
the hard spectral component above $\sim 10$\,keV
is in the shape of a valley bottomed out when flux peaks
during a shot,
which is  different from that in respect to the softer component.
(2) During an average shot of Cyg X-1 in the soft state, the hardness
in regard to a softer band lower $\sim 10$\,keV peaks when
the flux peaks .
(3) For the hard and transition states, shots are softer than
the time average emission, a sharp rise is appeared at about the
shot peak in the average profile of hardness ratio in respect to
a soft band.
(4) The correlation coefficients between the intensity and hardness
ratio on timescales between 0.01\,s and 50\,s are negative or near zero
when Cyg X-1 is in the hard or transition
states. In the soft state, the correlation coefficient in the hard
band decreases monotonically and reaches near $-1$
as the timescale increases from 0.01\,s to 50\,s; by contrast, the correlation
in regard to a soft band is positive and increases when timescale
increases.

The significant difference observed in Cyg X-1 of soft state
between the shot spectral evolution in the hard band above $\sim
10$\,keV and that in respect to a softer band indicates that
different mechanisms dominate the shot processes in the two energy
bands. It is broadly believed that the process of producing the
hard power-law like spectrum is thermal Comptonization: hard
emission above $\sim 10$\,keV is produced by inverse Compton
scattering of soft seed photons by hot electrons of temperature
$kT_e\le 100$\,keV (e.g. Sunyaev \& Tr\"{u}mper 1979). When a seed
photon of energy $h\nu_i$ collides with a relativistic electron in
the hot corona with Lorenz factor $\gamma$, the energy, $h\nu$, of
the emitted photon is given by $h\nu \approx \gamma^2h\nu_i$ (e.g.
Lang 1999). After a collision from an electron of energy 100\,keV,
the energy of a seed photon of 10\,keV will be lifted to $\sim 14$\,keV. 
Most scattered photons escaped from the hot corona undergo
only one collision and the hardness ratio in respect to the 13--16\,keV 
band decreases when the number of seed and scattered photons
increase. Thus the anti-correlation between the ratio of counts in
(16--60\,keV)/(13--16\,keV) and the hard band intensity during shots
in any spectral state can be understood by the Comptonization
process in the hot corona. Recently, by using a timescale
analysis technique of shot width, Feng, Li \& Zhang (2004) found
that the energy dependence of shortest width of X-ray shots from
the black hole binaries Cyg X-1, XTE J1550-564 and GRO J1655-40 in
their hard state in the energy band below about 10--20\,keV is
opposite to that in higher energy region: below about 10--20\,keV
the shortest width decreases with increasing energy, and above
about 10--20\,keV it increases with increasing energy. It is
expected that detailed study of shot spectral evolution and other
energy resolved timing in the hard band can help to understand the
hot corona and the shot production and propagation processes. 

 The evolution of shot hardness ratio in regard to a soft band below
$\sim 10$\,keV should depend on not only the hot gas, but also the
cold disk. It is natural to assume that shots are most probably
produced at the innermost region of the cold disk joined with the
hot corona which is a most turbulent region,
 and the steady component around a shot is a global
average of emission from different regions of the hot corona
(Li, Feng \& Chen 1999).
For the soft state, the disk may extend down to the last stable
orbit and the hot cloud is restricted to the corona surrounding
the disk \citep{esi98}. As the inner disk, and then the corona
embedding it, has a higher temperature, the shot spectrum
from the innermost region should be harder. For the hard and
transition states, the optically thick disk is truncated at
larger distance and jointed a spherical corona around the black hole.
Shots should be softer than the steady emission as they produce
and Comptonize at the outer part of the corona
with temperature lower than the average of the total corona.
The difference between the profile of hardness ration in regard to
a soft band in the soft state and that in the other states
may reflect the difference of their corona geometries.
The presence of an uprush of the hardness profile around the shot peak
in the hard and transition states is probably an evidence that
a shock wave is produced along with shot propagating in the spherical corona.

The correlation coefficients between the hardness ratio and intensity
obtained for different timescales and different states
of Cyg X-1 shown in Tables 2--3 and Figure 2 are qualitatively consistent
with the results from the shot analysis. The correlation analysis is
 made for the temporal variability in general, not only for shots.
The correlation being weaker at shorter timescales indicates
the existence of other uncorrelated rapid variations.
For the soft state, the correlation coefficients varying
(increasing or decreasing) monotonically and reaching a perfect
correlation (or anti-correlation) along with the timescale from 0.01\,s to 50\,s
may indicate that the effect of uncorrelated noise being weakened on larger
timescales.

The authors thank the referee for helpful comments and suggestions.
This work is supported by the Special Funds for Major State Basic Research
Projects and
the National Natural Science Foundation of China. The data analyzed in this
work are obtained through
the HEASARC on-line service provided by the NASA/GSFC.

\clearpage
\begin{figure}
\begin{center}
\epsscale{0.22}
\plotone{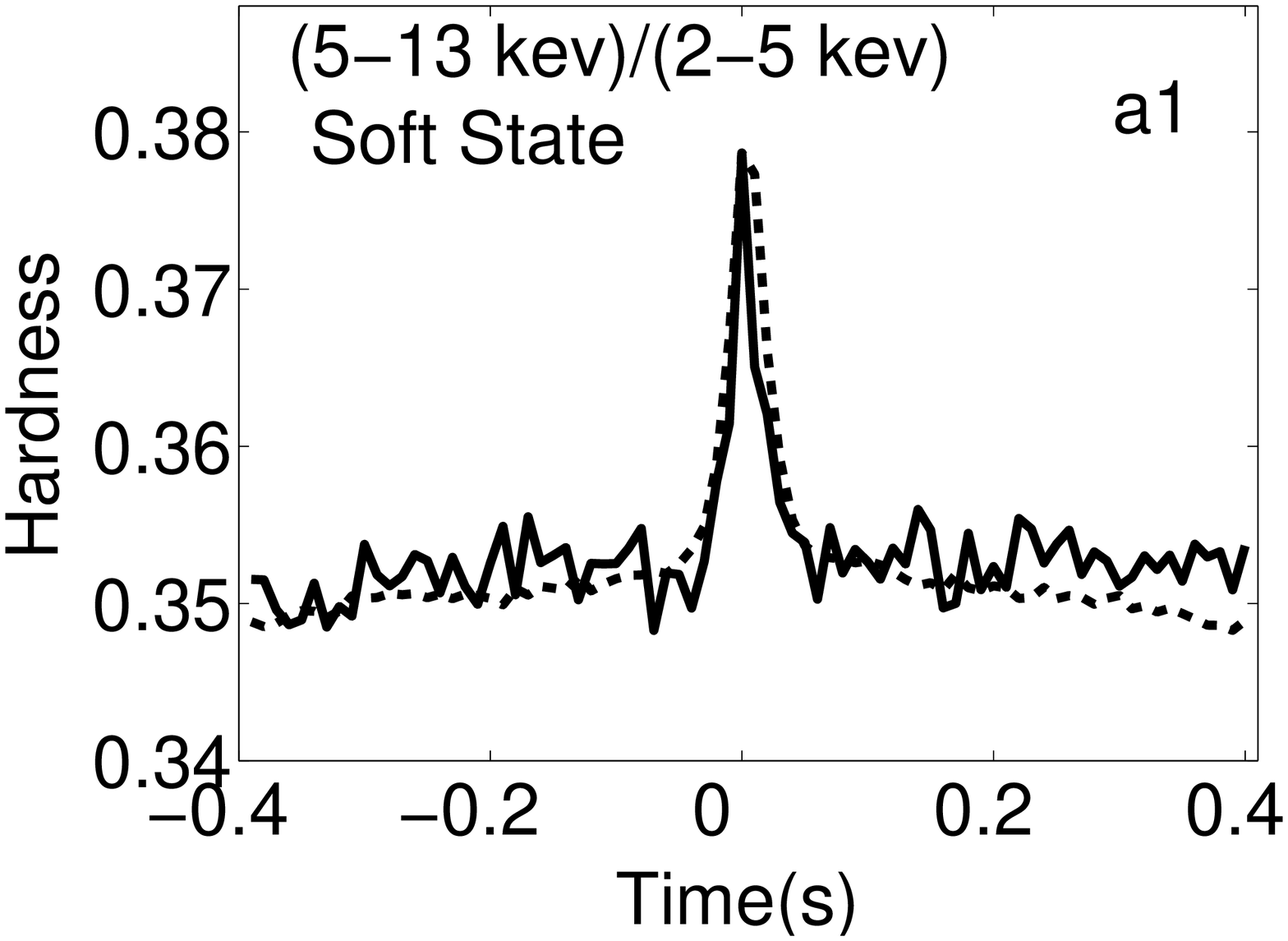}
\plotone{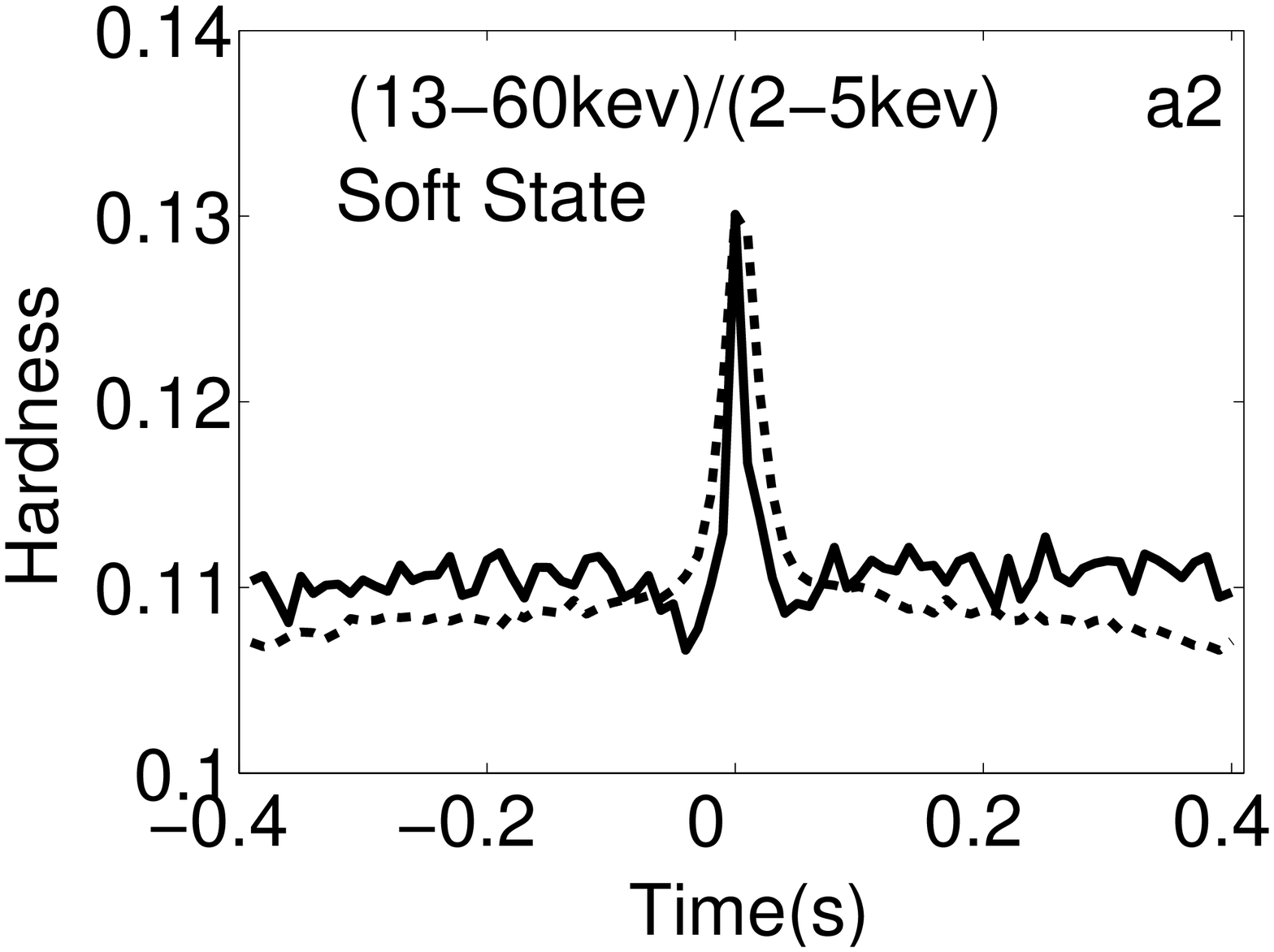}
\plotone{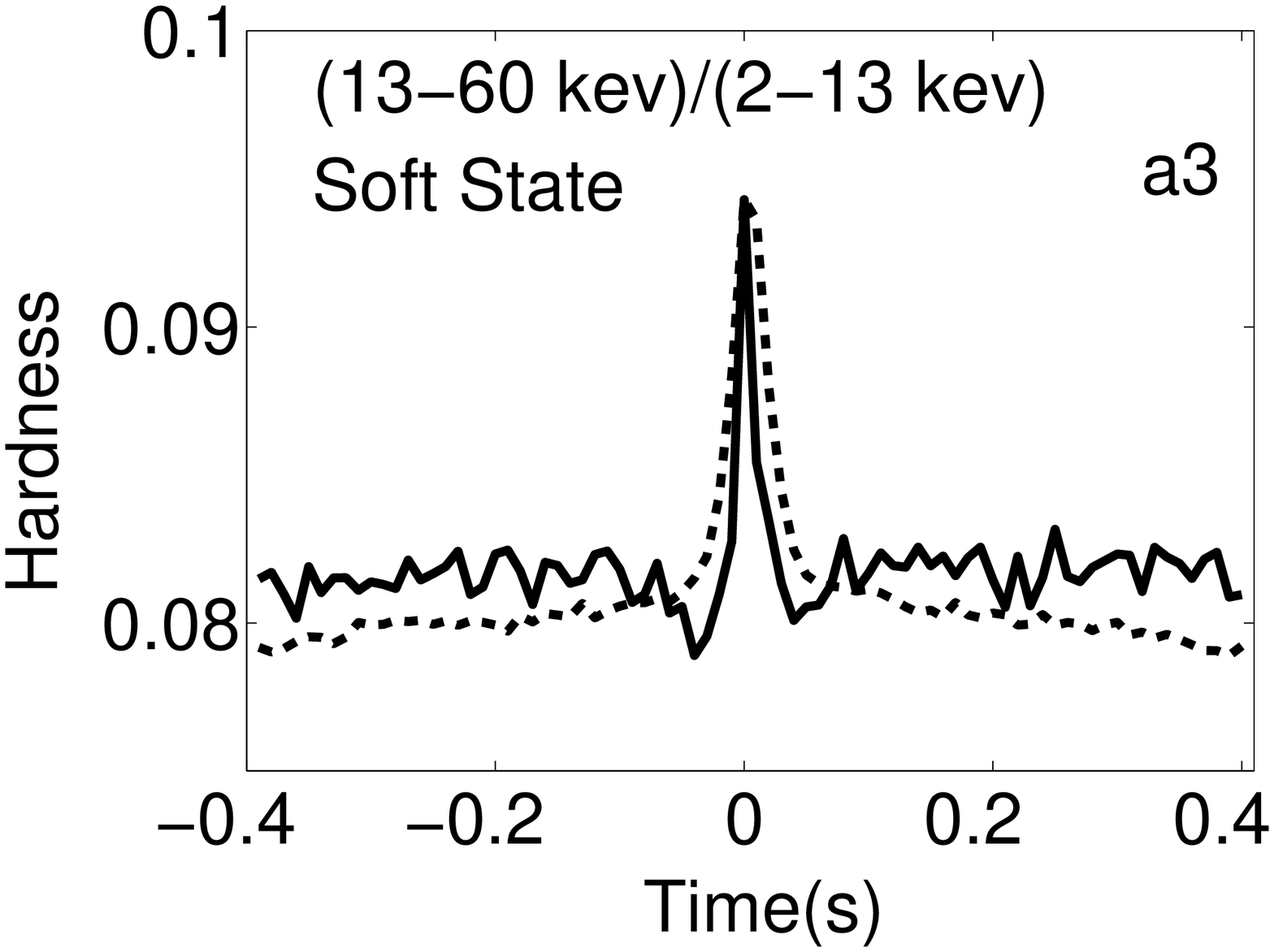}
\plotone{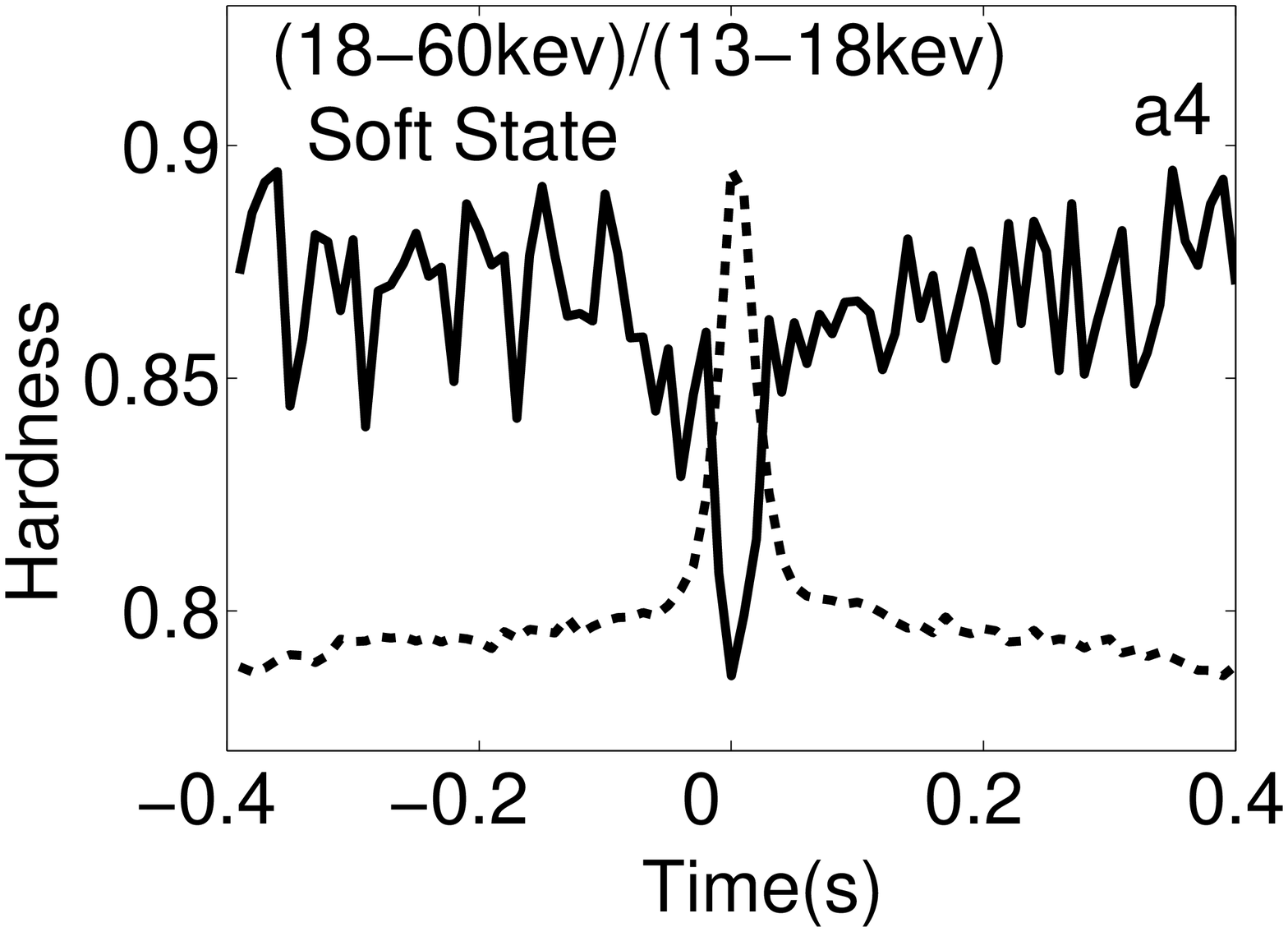}

\plotone{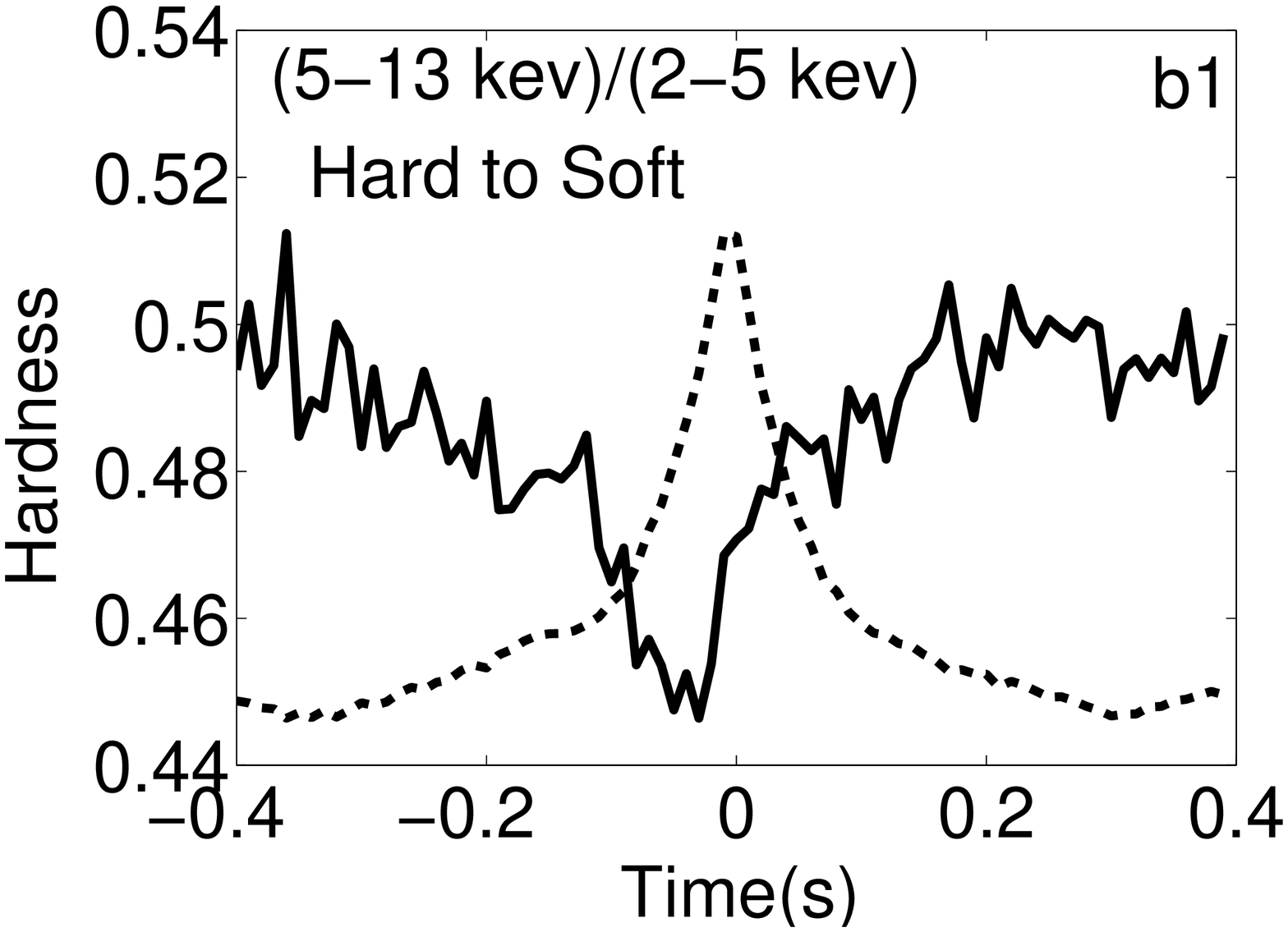}
\plotone{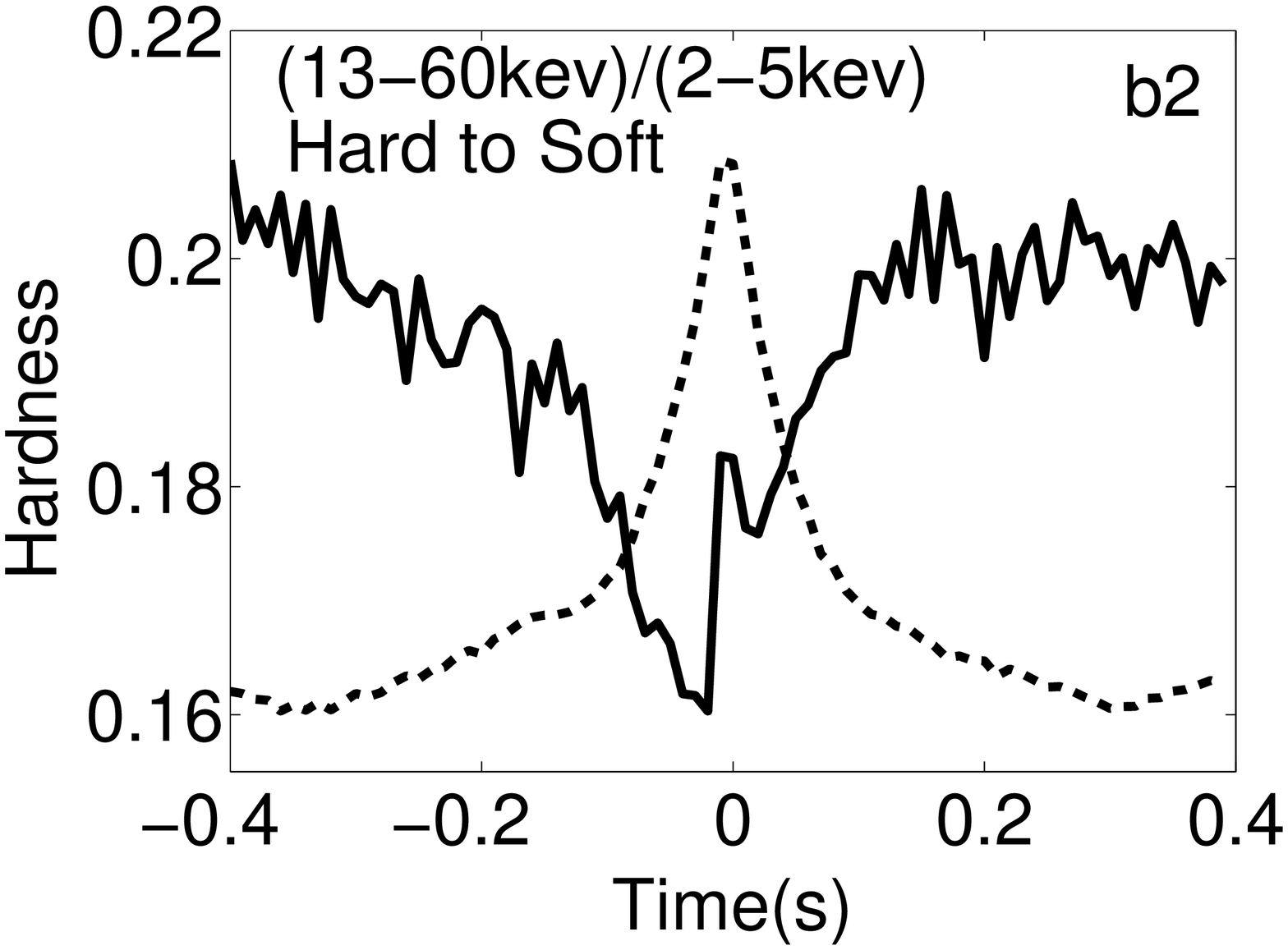}
\plotone{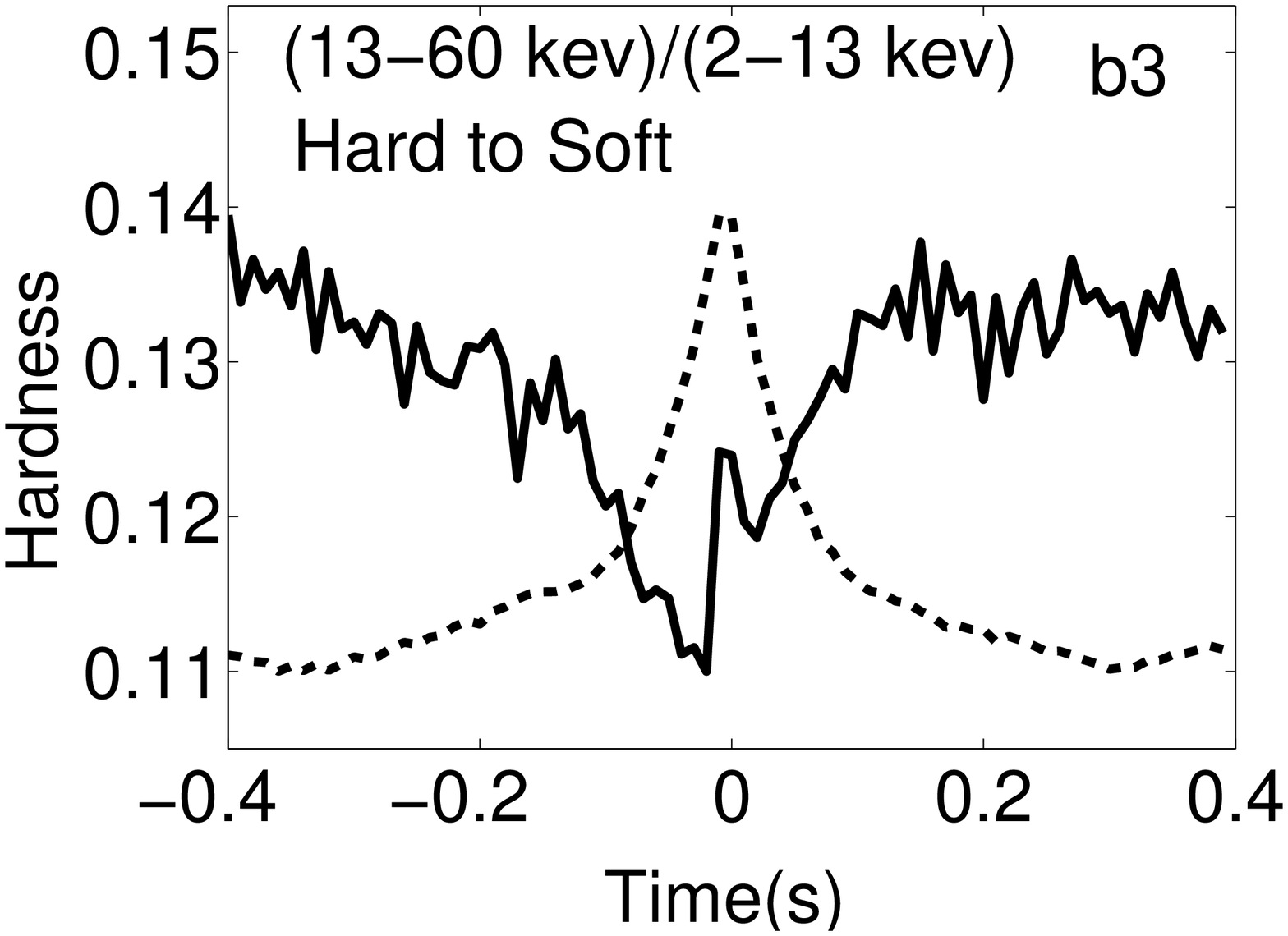}
\plotone{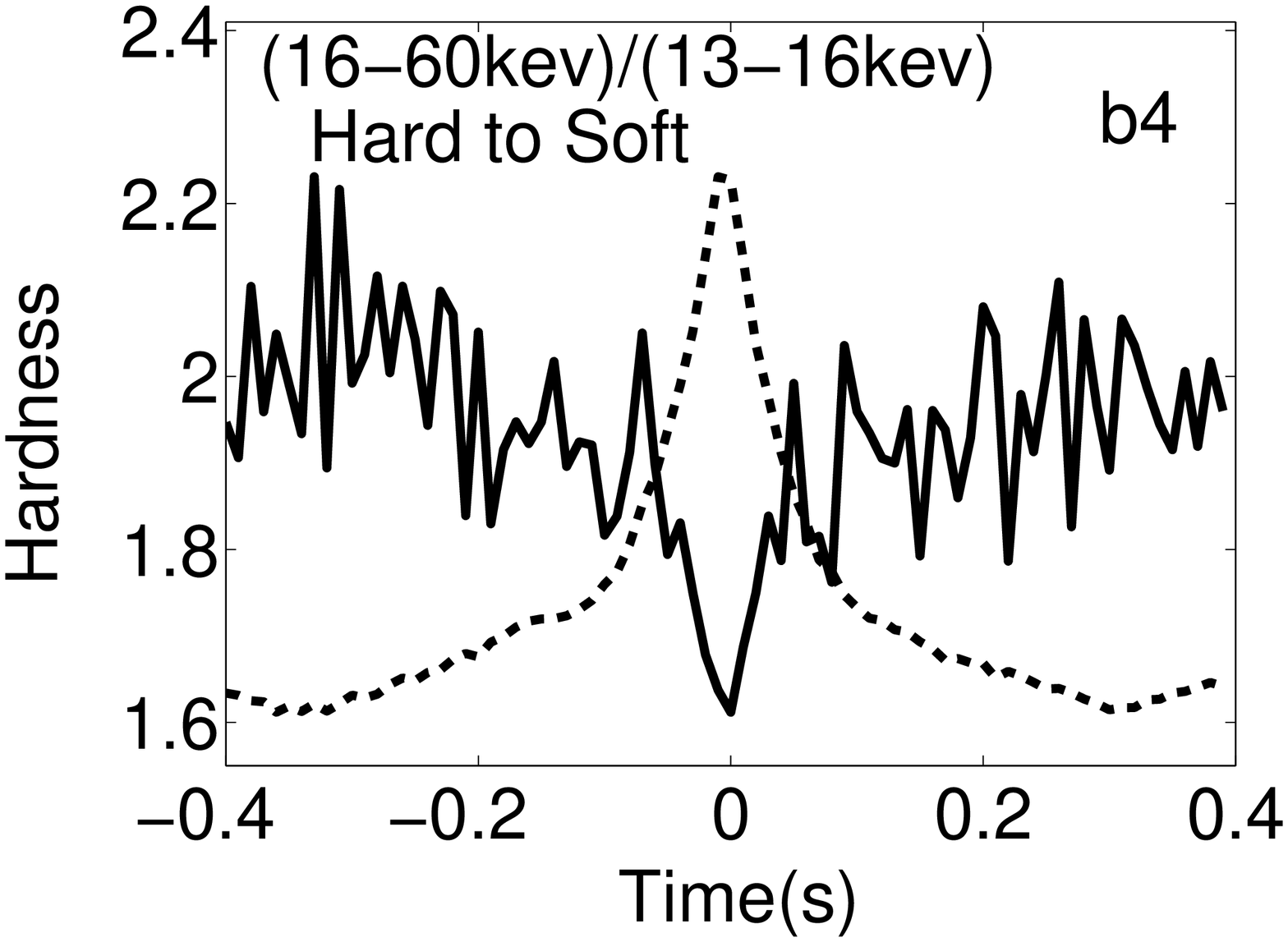}

\plotone{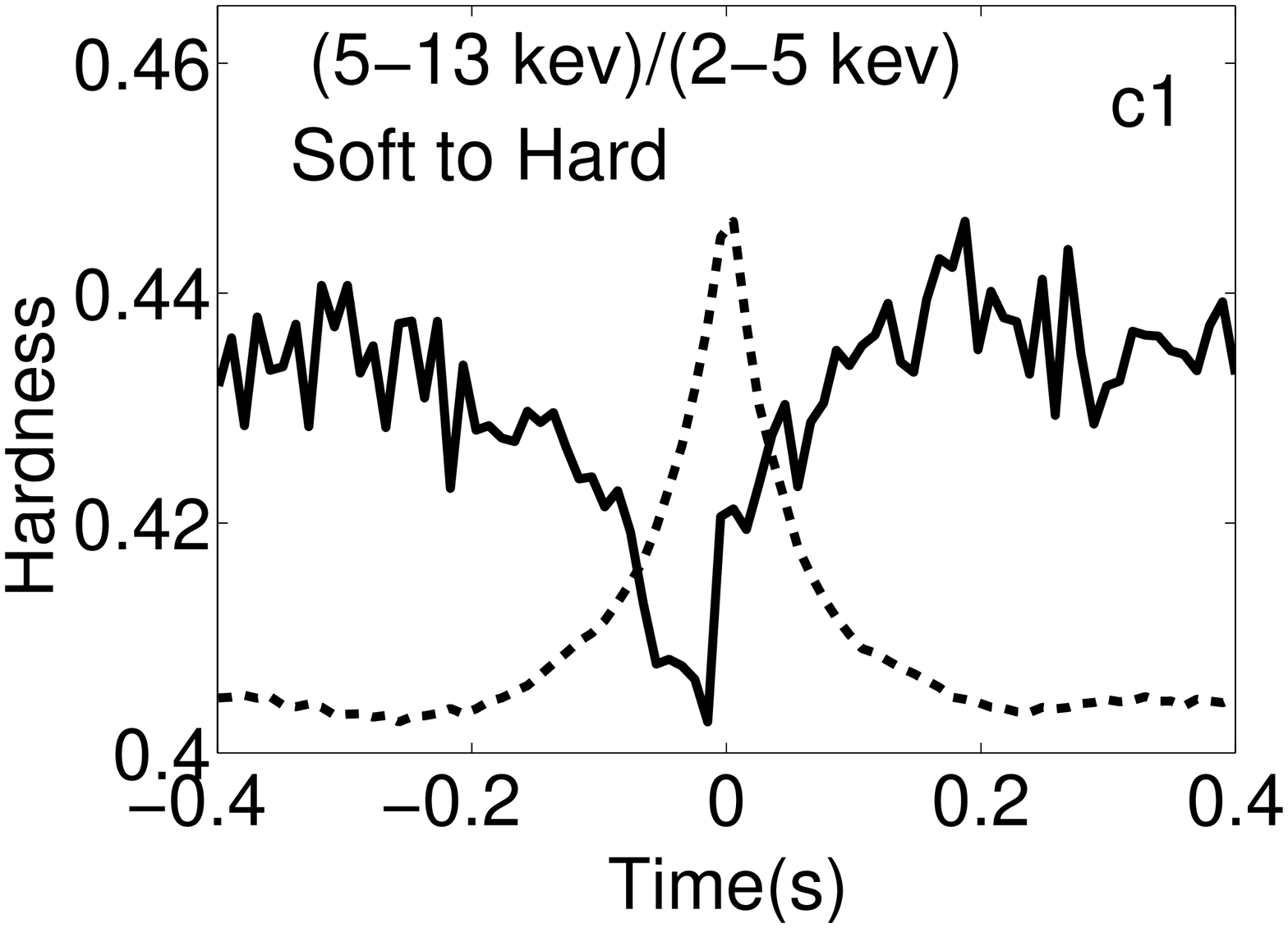}
\plotone{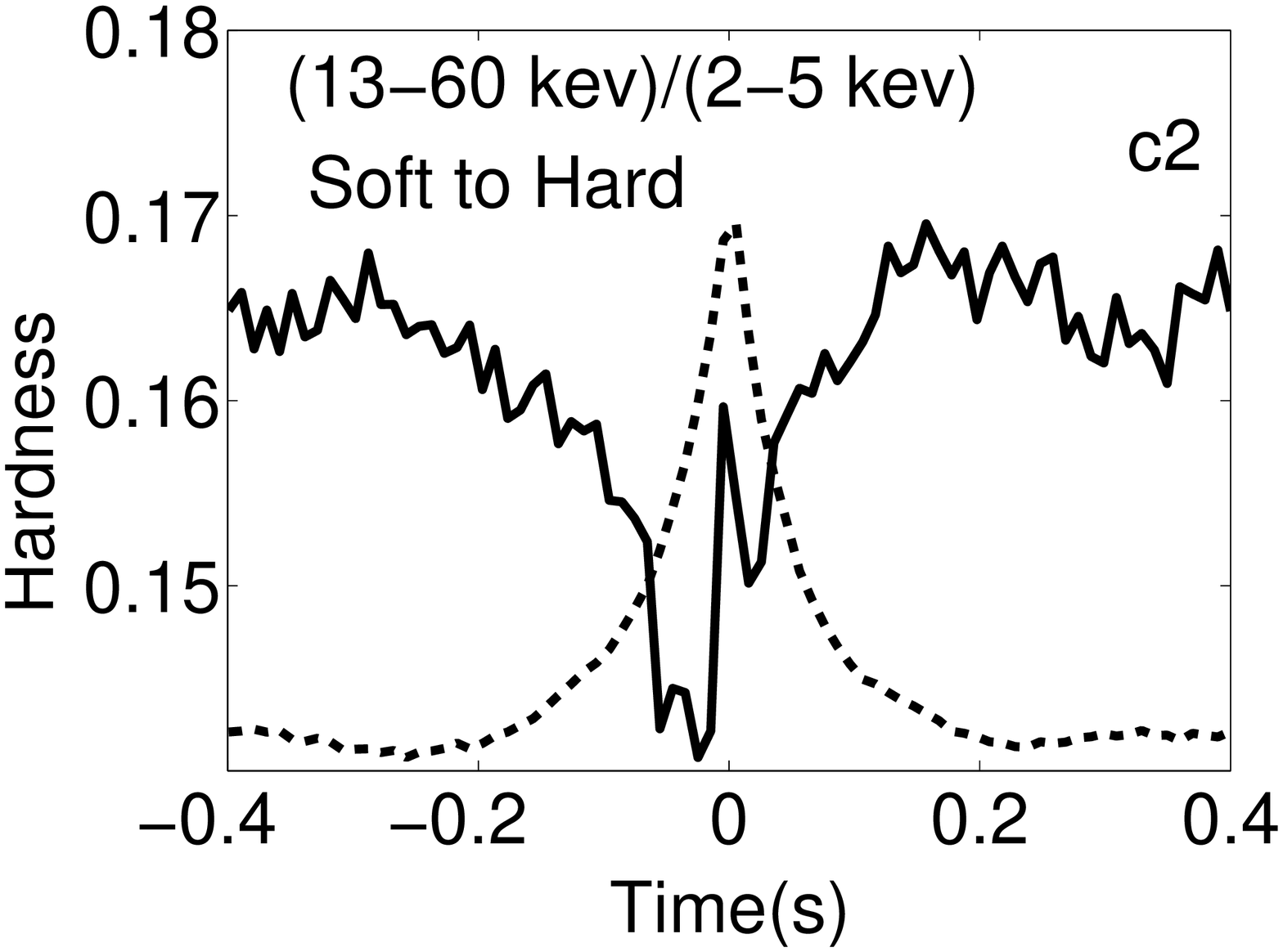}
\plotone{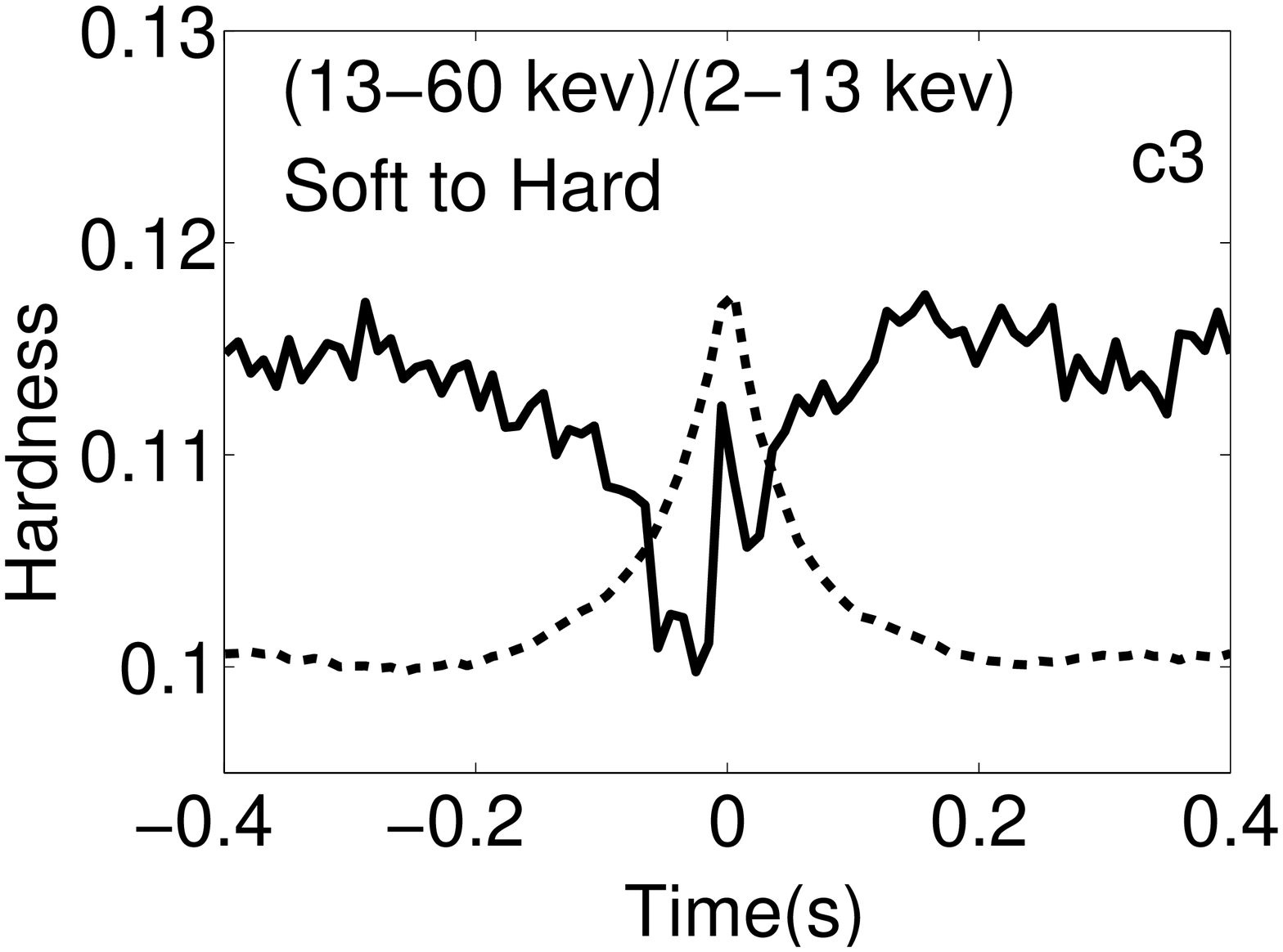}
\plotone{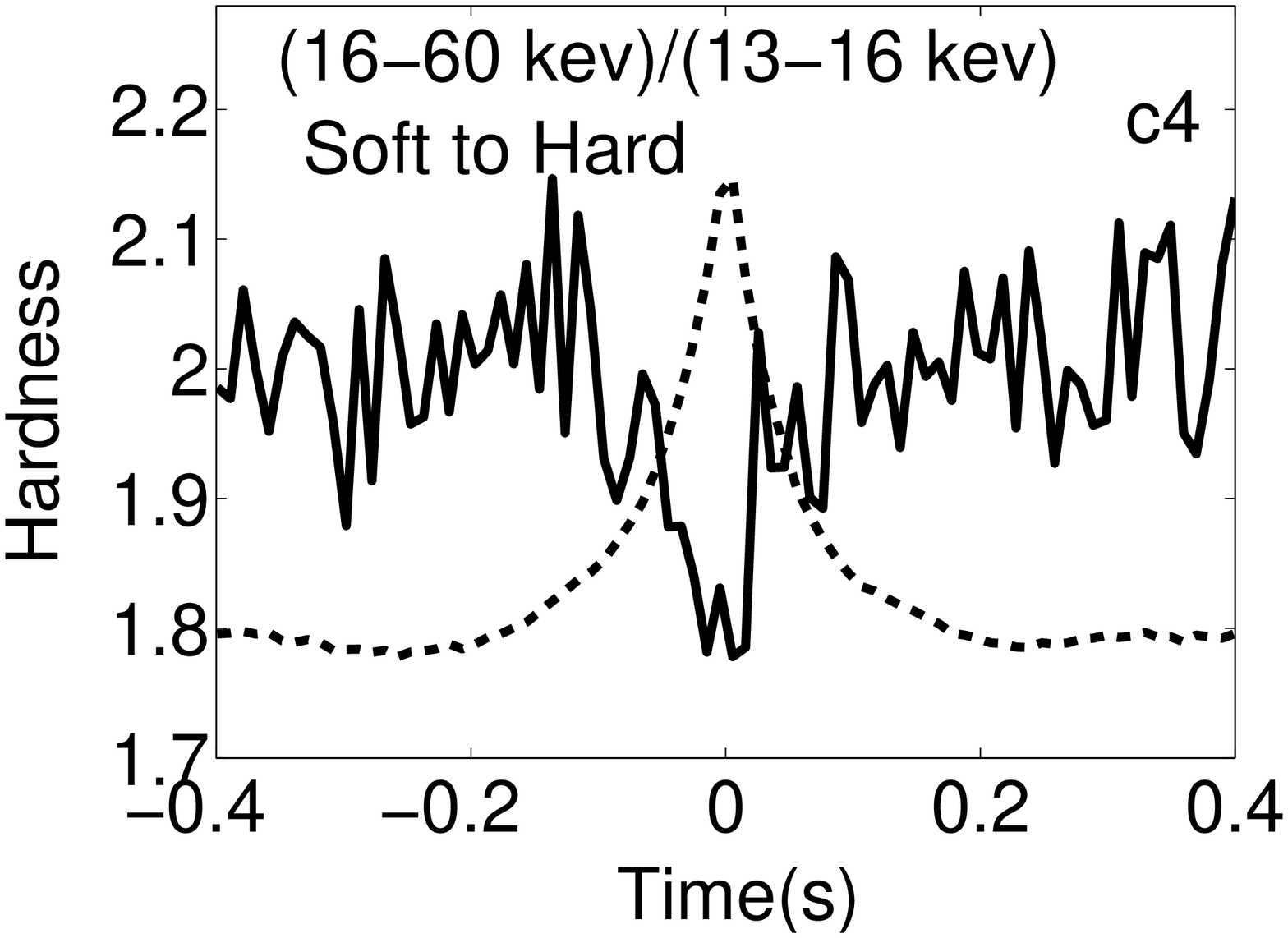}

\plotone{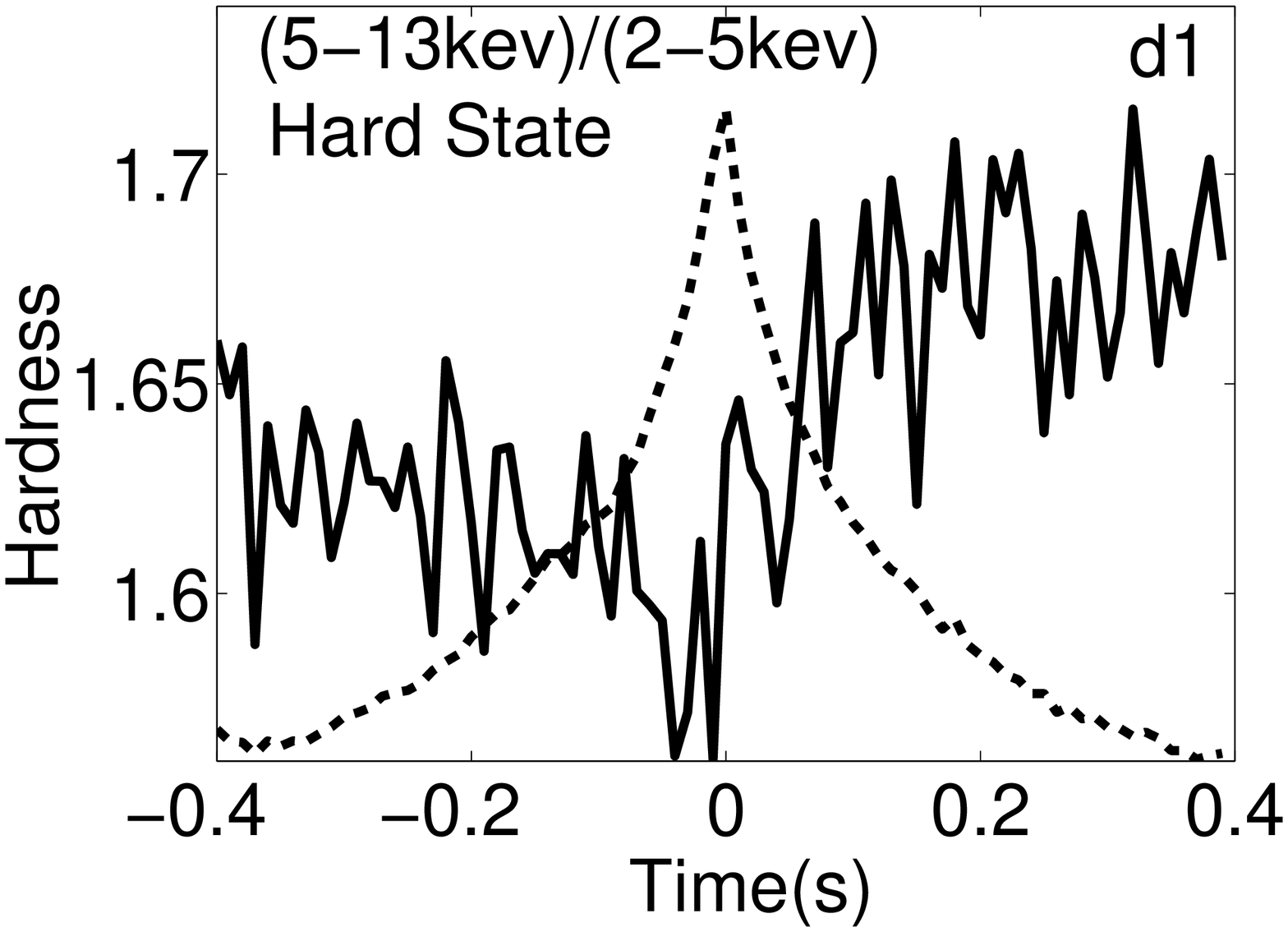}
\plotone{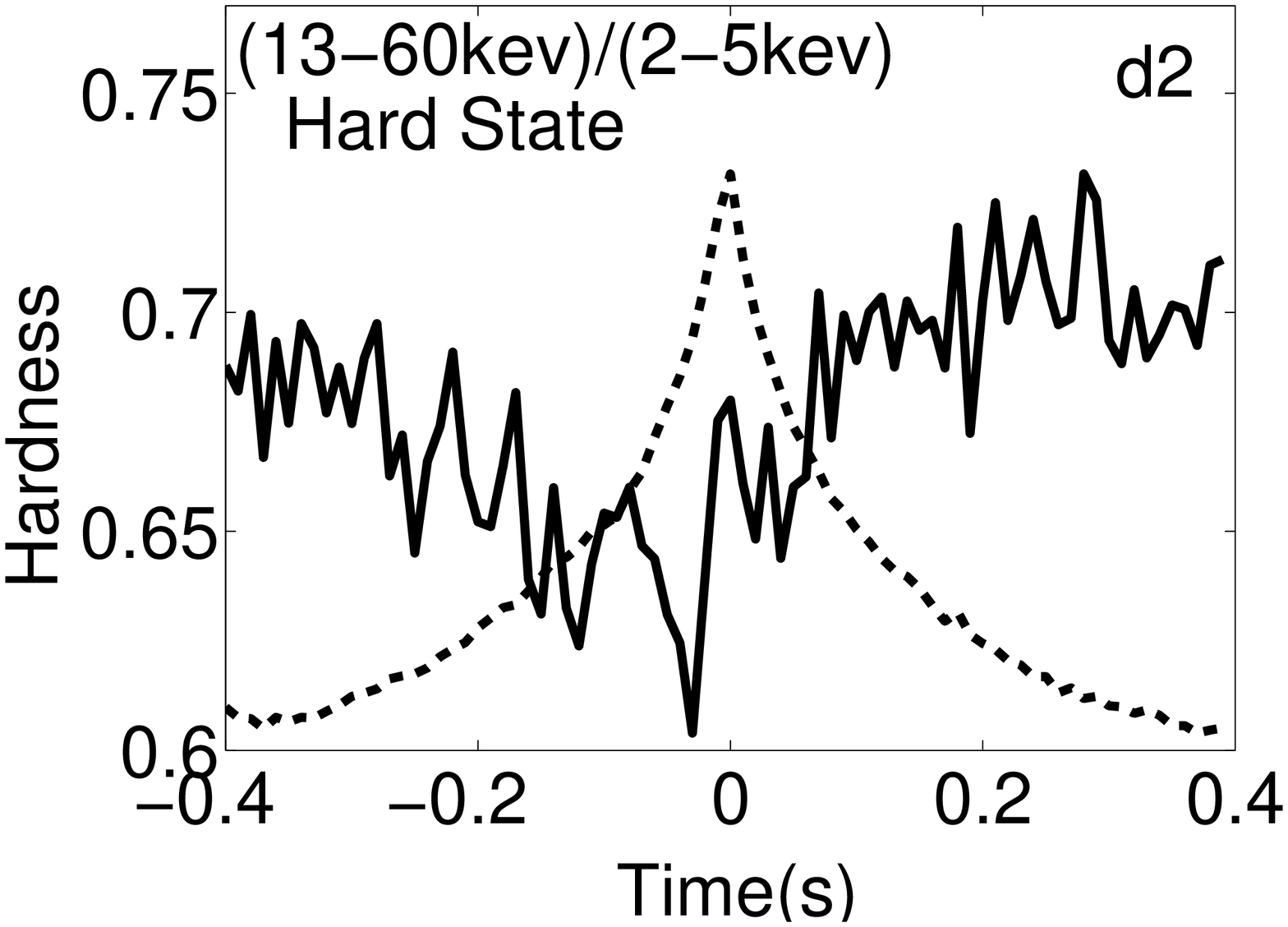}
\plotone{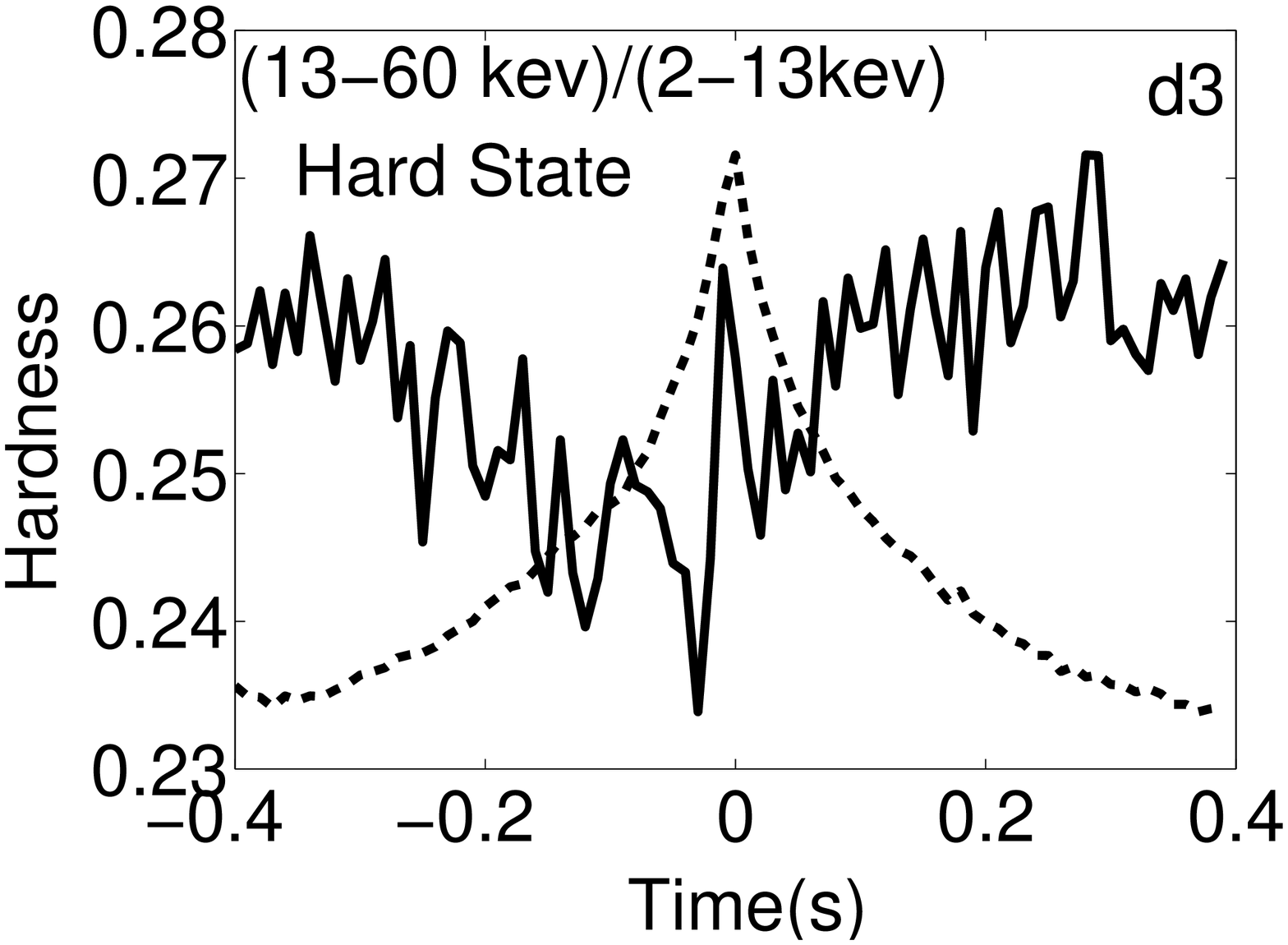}
\plotone{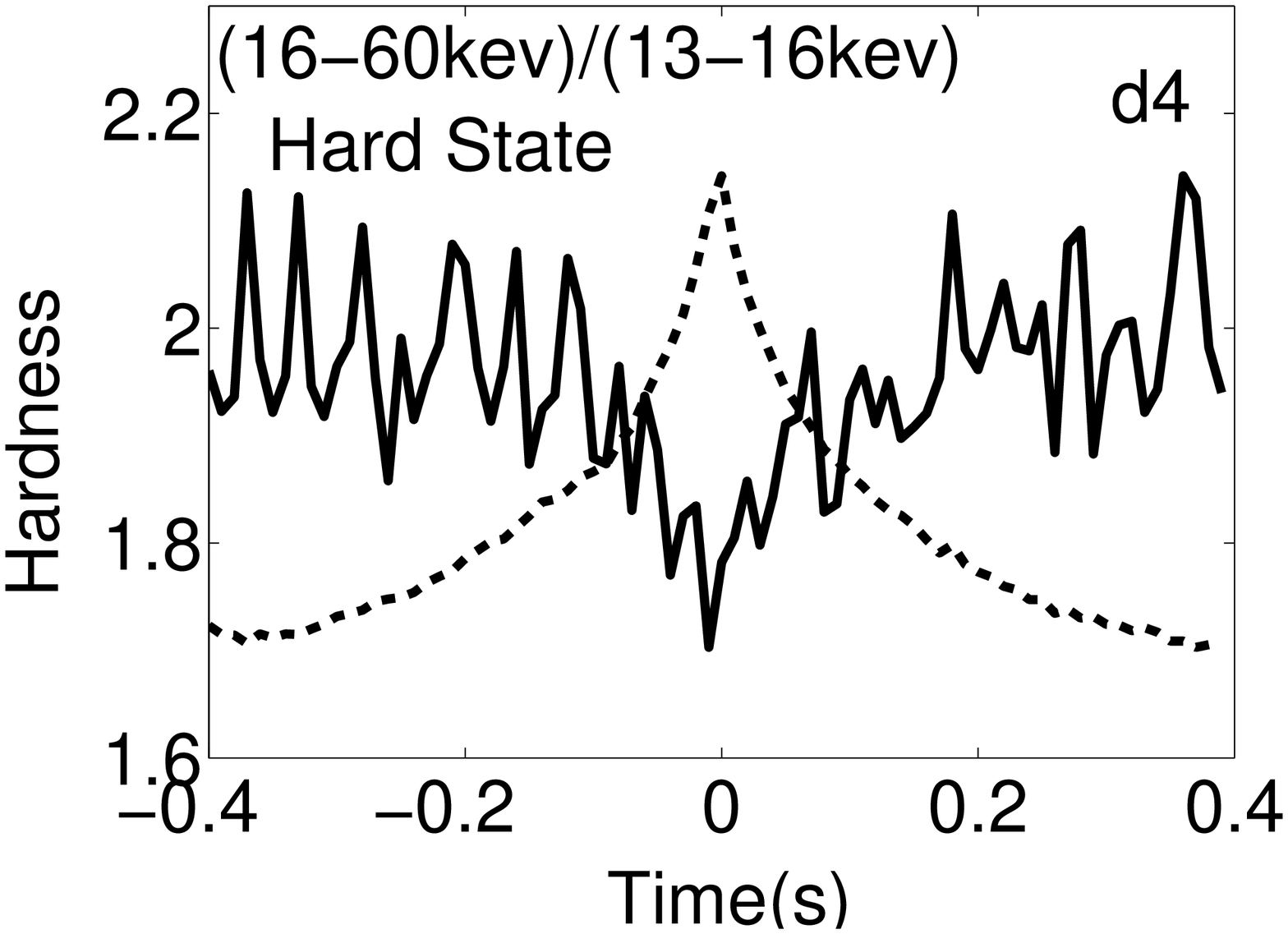} \caption{Flux and hardness ratio profiles of
average shots. {\it dashed line}: normalized shot flux profile.
{\it solid line}: hardness ratio profile. a1-a4: soft state;
b1-b4: hard to soft transition; c1-c4: soft to hard transition;
d1-d4: hard state. \label{fig1}}
\end{center}
\end{figure}

\clearpage
\begin{figure}
\begin{center}
\epsscale{0.42}
\plotone{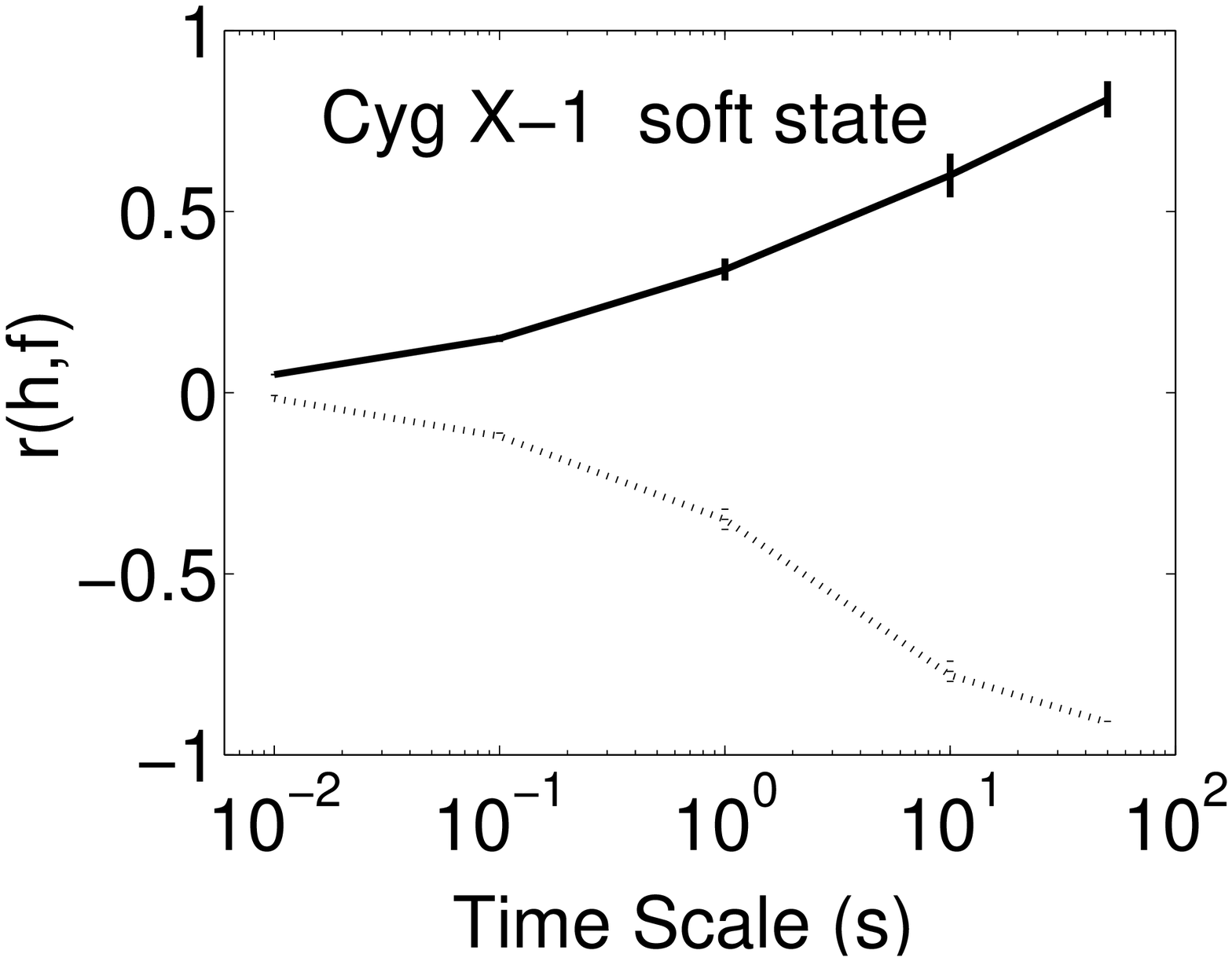}
\plotone{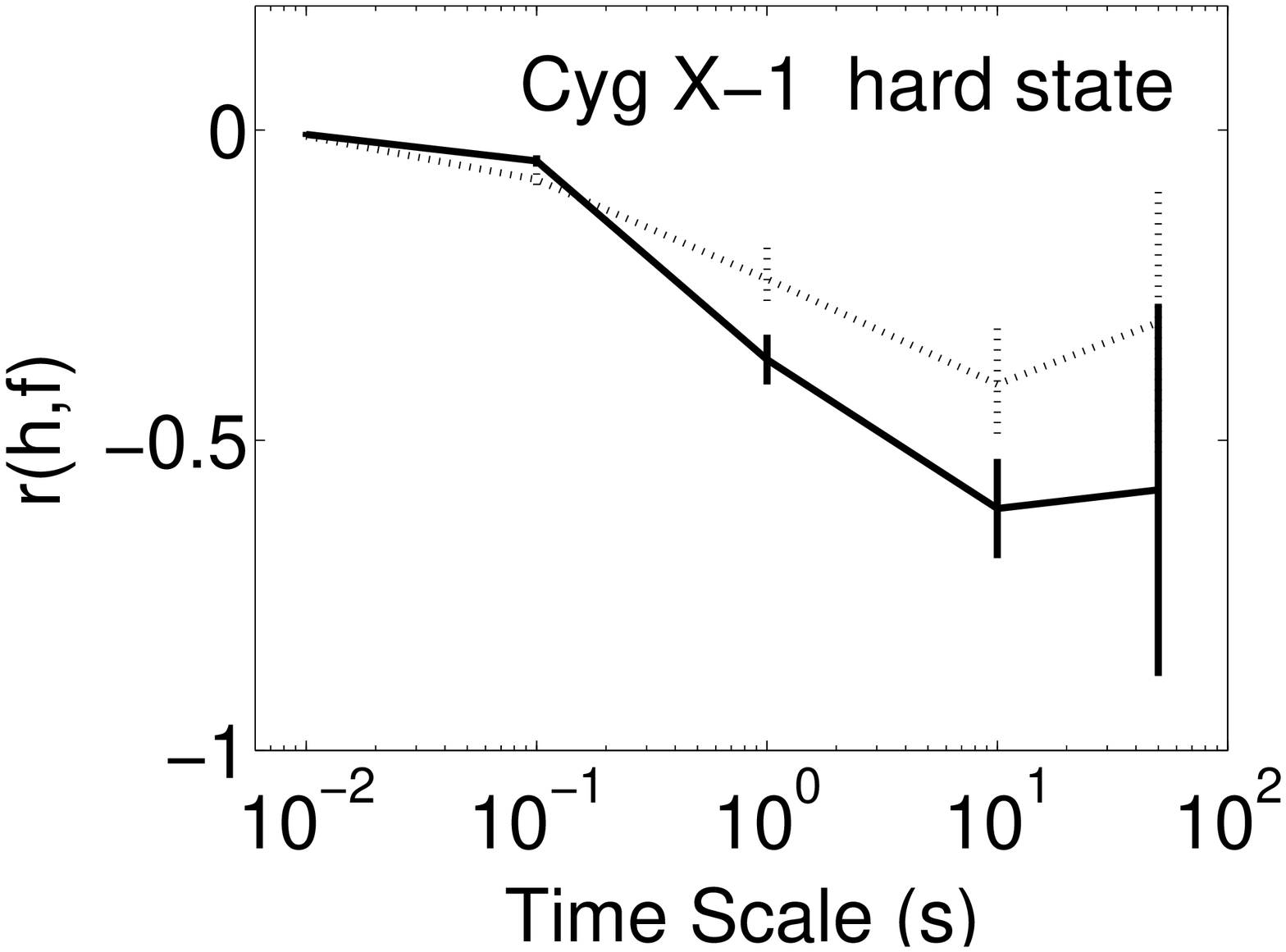}
\plotone{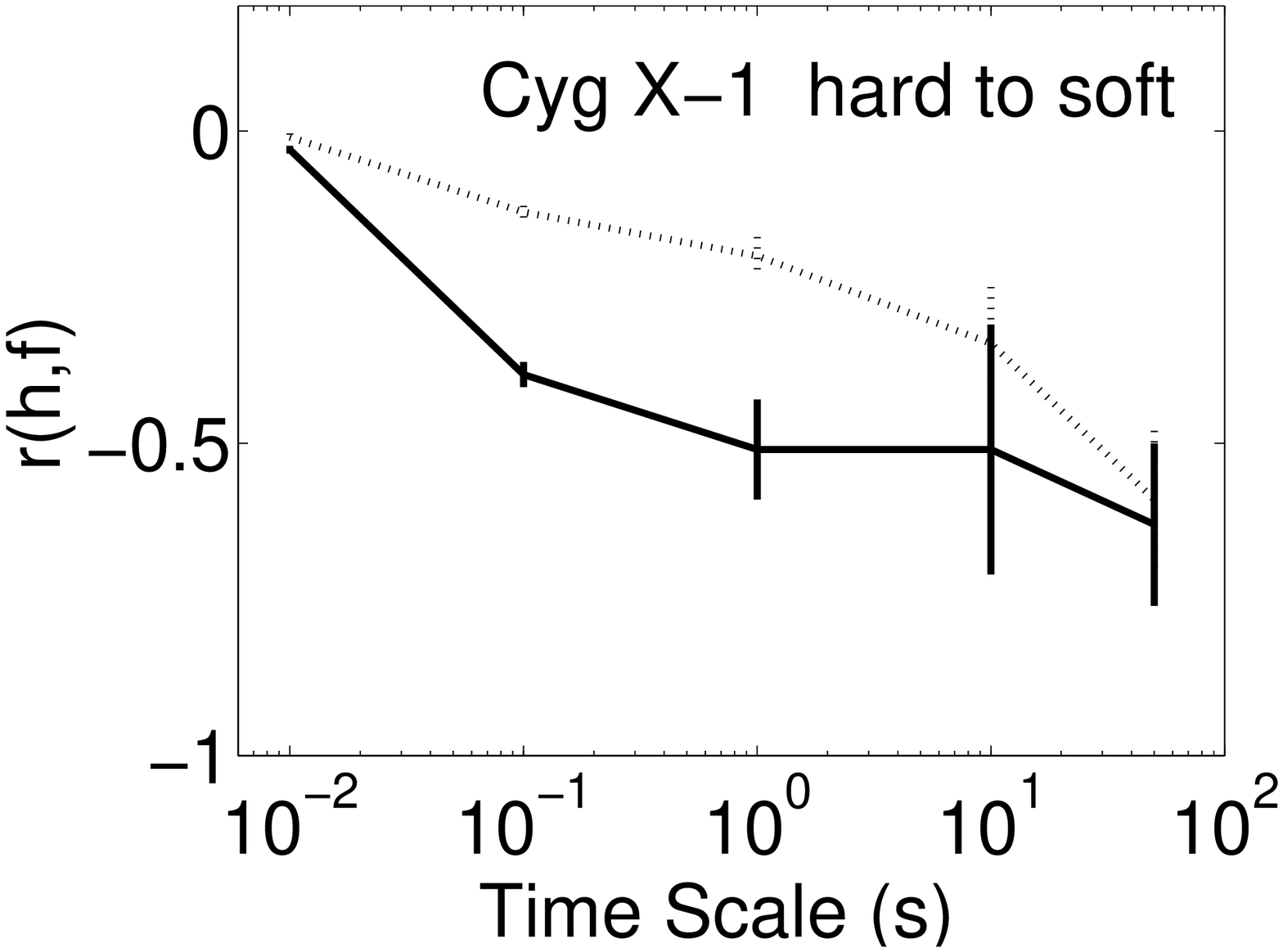}
\plotone{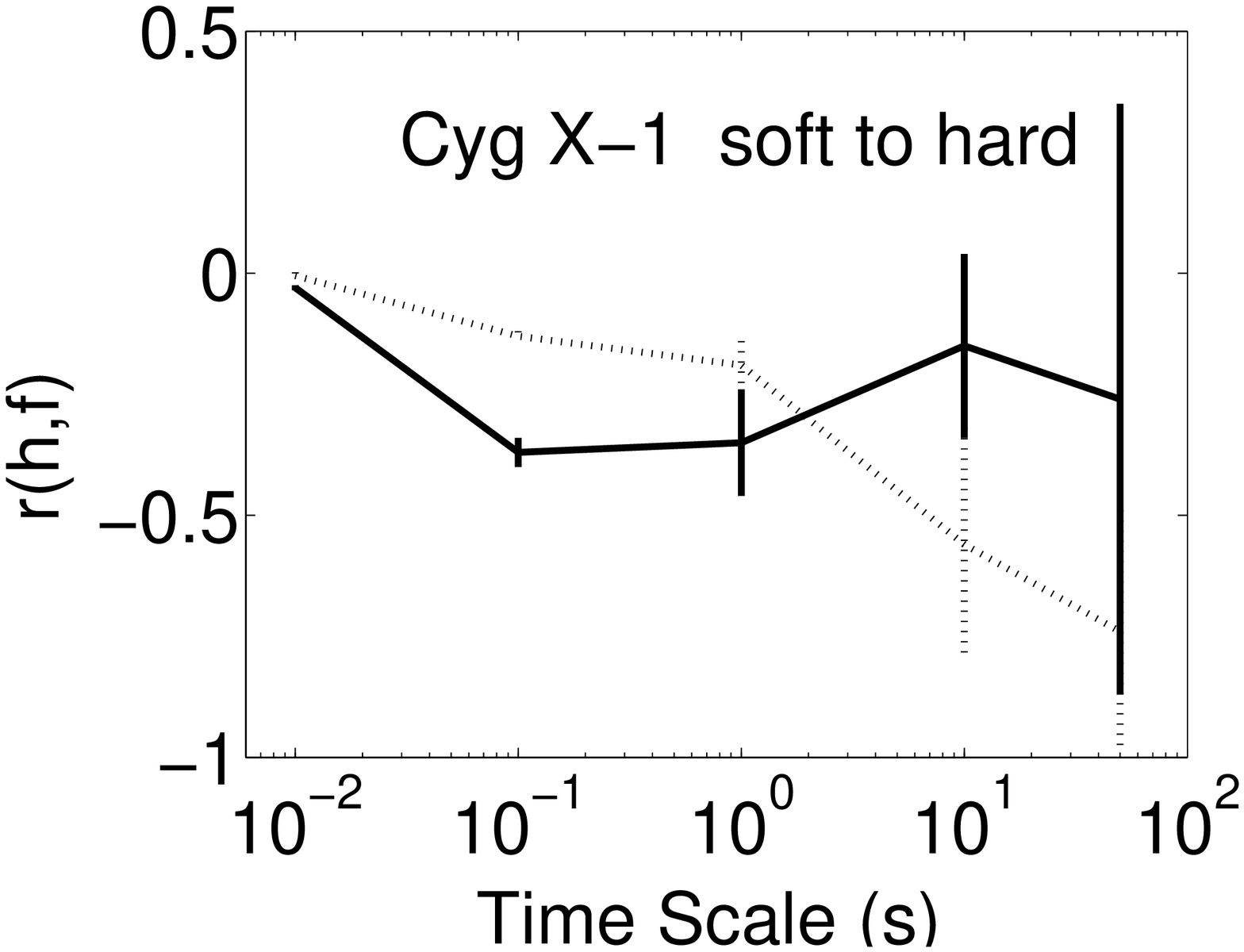} \caption{Correlation
coefficients between hardness and intensity vs. time scale. {\it
Solid line}: hardness in (13--60\,keV)/(2--5\,keV); {\it Dotted line}:
hardness in (16--60\,keV)/(13--16\,keV). {\it Upper-left panel}: soft
state; {\it Upper-right panel}: hard state; {\it Lower-left
panel}: hard to soft transition; {\it Lower-right panel}:soft to
hard transition. \label{fig2}}
\end{center}
\end{figure}

\clearpage

\begin{figure}
\begin{center}
\epsscale{0.44}
\plotone{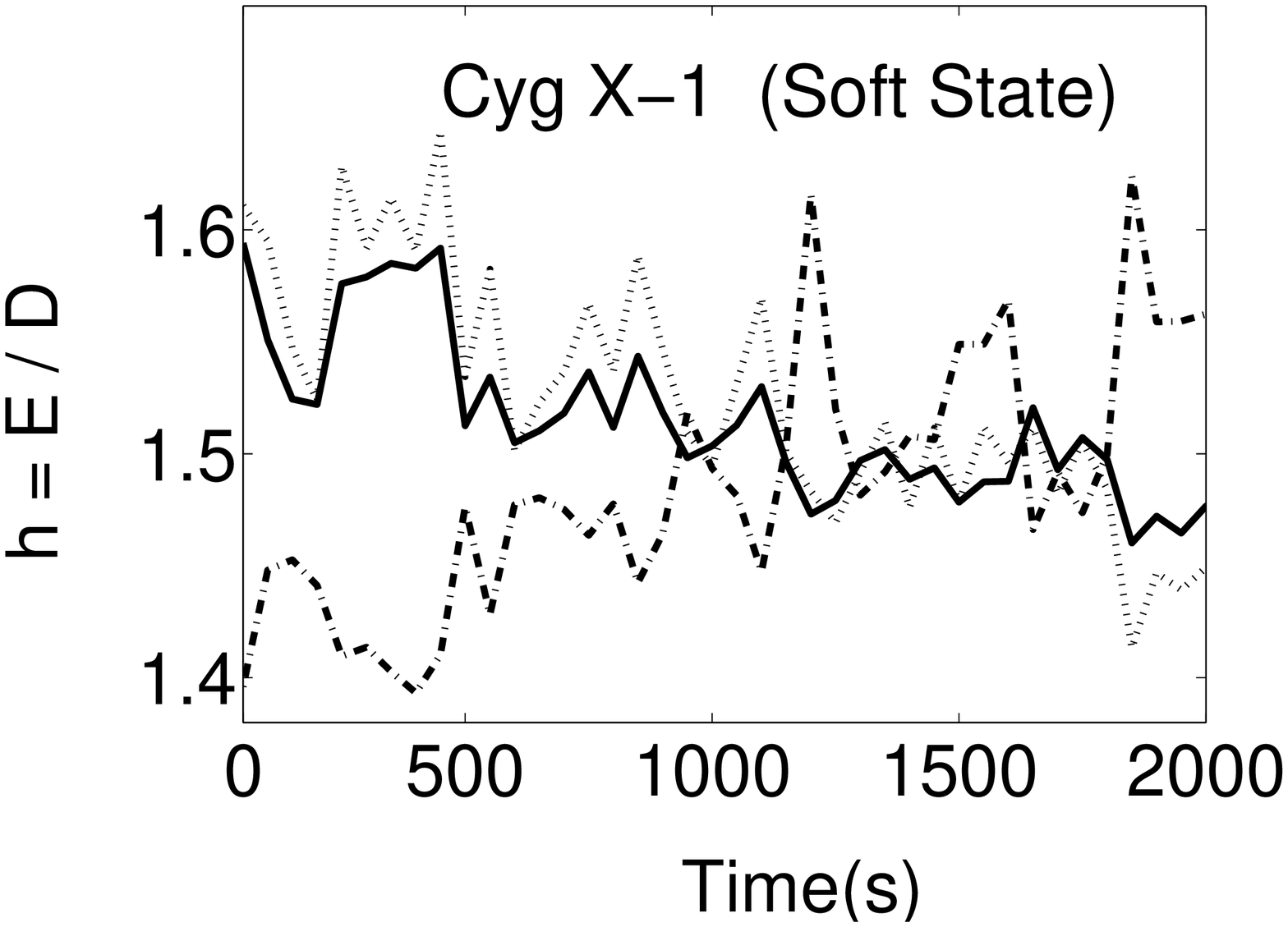} \epsscale{0.41}
\plotone{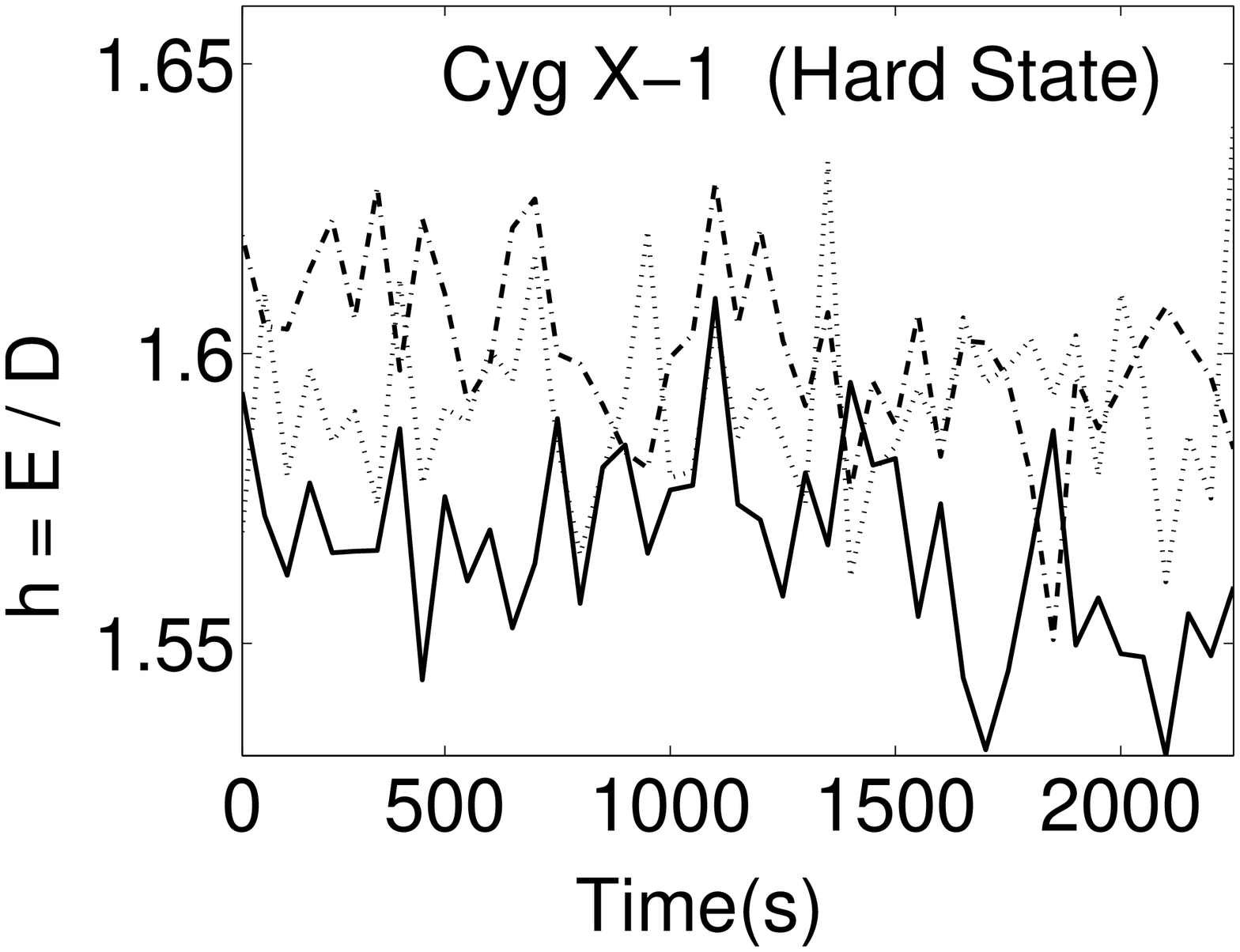} \caption{Intensity(E for (13--60\,keV), D for 
(13--16\,keV) in the figures) and hardness ratio profiles
in soft and hard states. {\sl Left panel}: soft state; {\sl
Right panel}: hard state. {\it Solid line}: 2--13\,keV light curve
in 50 s time bin in arbitrary unit. {\it dotted line}: hardness ratios in 
(13--60\,keV)/(2--6\,keV); {\it dash-dotted line}: hardness ratios in
(16--60\,keV)/(13--16\,keV).\label{fig3}}
 \end{center}
 \end{figure}

\clearpage

\begin{deluxetable}{cccc}
\tablecaption{List of RXTE observation of Cyg X-1 used for the
analysis. Dates are presented in DD/MM/YY format.\label{tbl-1}}
\tablewidth{0pt} \tablehead{ \colhead{State} & \colhead{Obs. ID} &
\colhead{Start Time} & \colhead{Stop Time}} \startdata
 Hard to Soft &10412--01--01--00 &23/05/96 14:14:05 &23/05/96 18:08:05 \\
 &10412--01--03--00 &30/05/96 07:47:05 &30/05/96 08:45:05\\
\cline{1-4}
Soft &10512--01--08--00 &17/06/96 07:59:05 &17/06/96 09:08:05\\
 &10512--01--08--02 &17/06/96 04:47:05 &17/06/96 05:44:05\\
\cline{1-4}
Soft to Hard&10412--01--05--00 &11/08/96 07:02:05 &11/08/96 08:25:05\\
 &10412--01--07--00 &12/08/96 14:41:05 &12/08/96 15:59:05\\
\cline{1-4}
Hard&10236--01--01--03 &17/12/96 12:45:42 &17/12/96 13:25:13\\
&10236--01--01--04 &17/12/96 22:21:42 &17/12/96 00:42:13\\
\enddata
\end{deluxetable}

\begin{deluxetable}{c|ccccc}
\tablecaption{Average Correlation Coefficients between Hardness
(13--60\,keV)/(2--13\,keV) and Intensity of 2--60\,keV in Cyg X-1}
\label{tbl-2}\tablewidth{0pt} \tablehead{\colhead{State} & &
&\colhead{ Time Scale}
 \cr
 &\colhead{0.01\,s} & \colhead{0.1\,s} &\colhead{1\,s}
 &\colhead{10\,s} &\colhead{50\,s} } \startdata
Soft & 0.05$\pm0.002$ & 0.15$\pm0.01$  &0.34$\pm0.03$&0.60$\pm0.06$&0.81$\pm0.05$  \\
Hard & -0.007$\pm0.004$ & -0.05$\pm0.009$ & -0.37$\pm0.04$&-0.61$\pm0.08$ &-0.58$\pm0.30$ \\
Hard to Soft & -0.03$\pm0.006$ &-0.39$\pm0.02$ & -0.51$\pm0.08$&-0.51$\pm0.20$&-0.63$\pm0.13$  \\
Soft to Hard& -0.03$\pm0.005$ & -0.37$\pm0.03$ & -0.35$\pm0.11$ & -0.15$\pm0.19$ &-0.26$\pm0.61$\\
\enddata
\end{deluxetable}

\begin{deluxetable}{c|ccccc}
\tablecaption{Average Correlation Coefficients between Hardness
(16--60\,keV)/(13--16\,keV) and Intensity of 13--60\,keV in Cyg X-1}
\label{tbl-3}\tablewidth{0pt} \tablehead{\colhead{State} & &
&\colhead{ Time Scale}
 \cr
 &\colhead{0.01\,s} & \colhead{0.1\,s} &\colhead{1\,s}
 &\colhead{10\,s} &\colhead{50\,s} } \startdata
Soft & -0.016$\pm0.01$ & -0.12$\pm0.01$  & -0.35$\pm0.03$&-0.78$\pm0.04$&-0.91$\pm0.004$  \\
Hard & -0.01$\pm0.003$ & -0.08$\pm0.01$ & -0.24$\pm0.05$&-0.41$\pm0.09$ &-0.31$\pm0.21$ \\
Hard to Soft & -0.01$\pm0.006$ &-0.13$\pm0.01$ & -0.20$\pm0.03$&-0.34$\pm0.09$&-0.59$\pm0.11$  \\
Soft to Hard& -0.005$\pm0.007$ & -0.13$\pm0.01$ & -0.19$\pm0.05$& -0.56$\pm0.24$&-0.74$\pm0.25$\\
\enddata
\end{deluxetable}

\end{document}